\documentclass[12pt]{article}

\usepackage{graphics,psfrag}
\usepackage{amsthm,amssymb,epsfig,amsmath,euscript,array,cite}
\usepackage{color}
\usepackage{graphicx}
\usepackage{hyperref}

\newcommand{\be}{\begin{equation}}
\newcommand{\ee}{\end{equation}}
\def\bea{\begin{eqnarray}}
\def\eea{\end{eqnarray}}
\newcommand{\eq}[1]{(\ref{#1})}
\newcommand{\la}[1]{\label{#1}}

\newcommand{\beq}{\begin{equation}}
\newcommand{\eeq}{\end{equation}}
\newcommand{\ben}{\begin{eqnarray}}
\newcommand{\een}{\end{eqnarray}}
\newcommand{\bes}{\begin{subequations}}
\newcommand{\ees}{\end{subequations}}
\newcommand{\blg}{\begin{align}}
\newcommand{\elg}{\end{align}}

\newcommand{\cN}{{\cal N}}

\newcommand{\prt}[1]{{\left( {#1} \right)}}

\newcommand{\startappendix}{
\setcounter{section}{0}
\renewcommand{\thesection}{\Alph{section}}}

\def\one{\mbox{1 \kern-.59em {\rm l}}}

\def\l({\left(}
\def\r({\right(}

\def\a{\alpha}

\def\d{\delta}

\def\l{\lambda} \def\L{\Lambda}
\def\m{\mu} \def\n{\nu}

\def\r{\rho}
\def\s{\sigma}  
\def\t{\tau}
\def\th{\theta}

 \def\cB{{\cal B}} 
  \def\cF{{\cal F}}
 \def\cH{{\cal H}} 
  
 \def\cN{{\cal N}} \def\cO{{\cal O}}

  \def\cX{{\cal X}}

\def\kt{\tilde{k}}

\begin{document}


\vspace{14pt}

\begin{center}

{\Large \bf
$k$-Strings as Fundamental Strings
}
\vspace{18pt}

{\bf
 Dimitrios Giataganas
 }

{\em
{} Department of Nuclear and Particle Physics,\\
Faculty of Physics, University of Athens,\\
Athens 15784, Greece
\vskip .15in
{}  Rudolf Peierls Centre for Theoretical Physics,\\
University of Oxford, 1 Keble Road,\\
Oxford OX1 3NP, United Kingdom
}
\vskip .1in
{\small \sffamily
dgiataganas@phys.uoa.gr
}\vskip .2in
\end{center}
\vskip .7in

\centerline{\bf Abstract}

It has been noticed that the $k$-string observables can be expressed in terms of the fundamental string ones. We identify a sufficient condition for a generic gravity dual background which when satisfied the mapping can be done. The condition is naturally related to a preserved quantity under the T-dualities acting on the Dp-brane describing the high representation Wilson loops. We also find the explicit relation between the observables of the heavy $k$-quark and the single quark states.
As an application to our generic study and motivated by the fact that the anisotropic theories satisfy our condition, we compute the width of the $k$-string in these theories to find that the logarithmic broadening is still present, but the total result is affected by the anisotropy of the space.

\setcounter{page}0
\newpage

\section{Introduction and Motivation}

The $k$-strings are combinations of quark-antiquark pairs where each quark it is separated from the antiquark by a distance $R_0$ and each pair separated by the other by a distance $\d$, with $R_0\gg\d$. The common flux tube formed between the pairs is the $k$-string. In confining theories the potential in the tube takes the form
\be
V=\s_k R_0+\frac{c_1}{R_0}
\ee
where $\s_k$ is the string tension and $c_1$ is a dimensionless constant in the L\"{u}scher term. Both terms in the potential have been computed in several quantum field theories using lattice, field theory or gauge/gravity duality methods. Having $k$ number of strings that do not interact to each other when the pairs are kept well away, the string tension of the system is proportional to $k \s_1$ where $\s_1$ is the tension of each string. In this case we do not have a single bound state. By reducing the distance $\d$ between the pairs and bringing them closer, the gluonic made strings start interacting to make a flux tube, a bound state with string tension $\s_k$.

One may ask then the question what is the tension  $\s_k$ of the flux tube and if it is related to the tension of each string $\s_1$.
The answer is important since the $k$-strings are bound states that probe in a way the interaction of the gluonic strings to each other and their study expected to give a better understanding of the dynamics of confinement. There are two possible proposals for the form of the string tension, the so called Casimir scaling and the sine formula. The Casimir scaling was originally found to be exact in a 1+1 dimensional pure Yang-Mills theory and the transverse dimensions are absent \cite{casimir1}. It has been also obtained from the stochastic vacuum analysis \cite{casimir2}, as well as from a kind of dimensional reduction of the YM vacuum wavefunctional \cite{casimir3}. More particularly the string tension before the screening can be representation dependent and proportional to the quadratic Casimir of the theory
\be
\s_k=\frac{C_R}{C_{Fund}}\s_1,\quad\mbox{with}\quad T^{a} T^{a}= C_R I_R~,
\ee
where $C_R$ and $C_{Fund}$ is the quadratic Casimir in R representation and the fundamental respectively. The matrices $T^a$ are the $SU(N)$ generators in the representation $R$ of the theory and the $I_R$ is the unit matrix. For the antisymmetric representation the k-string tension
becomes
\be
\s_k=k\left(1-\frac{k-1}{N-1}\right)\s_f~
\ee
and its expansion in large $N$ leads to
\be
\s_k=k\left(1-\frac{k-1}{N}+ \cO\left(\frac{1}{N^2}\right) \right)
\ee
with higher order terms being increasing powers of $1/N$. 
The Casimir scaling was found in certain supersymmetric theories \cite{casimirsusy} as well as in a gauge-adjoint Higgs model in 2+1 dimension \cite{casimirhiggs}.

The alternative sine formula for the k-string tension has been
found in a confining softly broken $\cN=2$ supersymmetric Yang-Mills \cite{sine} and is representation independent. The sine formula takes the form
\be
\s_k=\frac{\sin\frac{\pi k}{N}}{\sin\frac{\pi}{N}}\s_1~,
\ee
and the large $N$ expansion becomes
\be
\s_k= k\left(1-\frac{\pi^2 \prt{k^2-1}}{6 N^2}+\cO\left(\frac{1}{N^4}\right)\right)\s_1~,
\ee
where the higher order terms are of order $1/N^2$. The sine formula has been also obtained in M theory fivebrane version of QCD \cite{sinemqcd}.

There is a debate on which formula describes better the k-string tension and whether or not any of them is accurate. The answer to this question is still considered as open, and there are several findings from string theory, lattice theory and field theory that favor one or the other formula. The answer may also be that the exact function of the string tension is none of these two proposed, although several other suggestions have been excluded \cite{Wingate:2000bb,Lucini:2004my}.

The two formulas have some essential differences. In the large $N$ expansion  the Casimir goes as $1/N$, while the sine as $1/N^2$. By considering the exchange of the gluons between the gluonic strings, which should be of minimum two since the color can not change on them, we can argue that the effect is of order $1/N^2$. This is true even when the quark-gluon interaction is taken into account, which  are non-planar. The possible $1/N$ terms appear only in the division to irreducible representations
and in general, the quadratic Casimir operator takes the following form:
\be\label{casim}
C_R\propto (\mbox{terms depend on rows and columns of the YT})-\frac{k^2}{N}~.
\ee
Assuming  the string tension to be proportional to the Casimir operator, its expansion will be of order $1/N$. The rest of the terms of \eq{casim} should not affect the tension since the result should be representation independent and therefore the final result will be of order $1/N$ in contrast to the expectation. This is an arguing in favor of the sine formula \cite{armoni}. Moreover, thinking the interaction between the world-sheet it appears to be of order $1/N^2$ terms. As well the large $N$ expansion in the field theory has the generic form of even powers of $1/N$. The above facts favor the sine formula with the $1/N^2$ expansion.

On the other hand recent and accurate lattice data results in 2+1 dimensions show better fitting of the Casimir formula \cite{teper1} for the k-string tensions.
On top of that, one should notice that the arguments of the previous paragraphs may favor the sine formula but seem not enough
to rule out the Casimir formula. It is possible in certain theories that the Casimir odd powers of $1/N$ terms cancel each other during the division to the irreducible representations, leaving the final result with only $1/N^2$ terms. This has been shown explicitly using the heat kernel action in Euclidean lattice gauge theory \cite{n1}. There at strong coupling the Wilson loop operators are known, and the string tensions are proportional to the quadratic Casimir. In the stage of
the decomposition of the fundamental representations into sum of the irreducible ones the odd powers of $1/N$ cancel each other because the contributions of the row-column conjugate Young tableaux (YT) \footnote{Tableaux that can be obtained from each other by interchanging the rows with the columns.}, give
exactly the opposite $1/N$ terms, while the self conjugate YT give only $1/N^2$ terms. Therefore, the large $N$ expansion of the field theory where only even powers of $1/N$ appear, can be consistent
with the Casimir formula scaling in the string tension.

Another notice on these computations is the arguing of the authors of \cite{n1}  that the limits where the time side of the relevant Wilson loops is taken large, and the large $N$ limit may not commute each other. The particular Wilson loops are of rectangular shape with one side being the time $T$ and the other the separation between the quarks $R$, with $T\gg R$.
The decomposition on the Wilson loop in energy eigenstates and the relevant analysis is sensitive to the order of the large $N$ and $T$ limits and the way that the limits are taken may affect the powers of $1/N$ expansion.  Summarizing the above discussion it seems that the Casimir formula can not be ruled out using only the large $N$ counting.

Having convinced ourselves that the these multiquark bound states are worthy studying and have important unanswered questions, we examine their connection to the single strings using the gauge/gravity duality.  In order to capture the interaction between the strings is not enough to use the Nambu-Goto action, because then the strings do not interact and the total string tension will be $k \s_1$. We consider appropriate probe Dp-branes with electric flux in the world-volume. These extend in two dimensions in the field theory space-time, since they represent heavy quarks and antiquarks and the rest of their dimensions wrap the internal space. It has been shown explicitly that in $AdS_5\times S^5$ the single string Wilson loops are appropriate for the fundamental representation \cite{wlf}, while for higher symmetric and antisymmetric representations are appropriate the D3 branes and the D5 branes \cite{Gomis:2006sb,wld5}  with electric flux respectively.

There are several studies of $k$-strings in the context of gauge/gravity dualities. In \cite{Callan:1999zf} the flux tube tension was computed in $2+1$ and $3+1$ confining setups, finding dependence on the number of dimensions. In Klebanov-Strassler theory \cite{ks} it has been found that the string tension follows approximate the sine
formula \cite{ksk}. In Cvetic, Gibbons, L\"{u}, Pope background \cite{cglp} the string tension has been found to take lower values than both Casimir and sine laws \cite{cglpk}, while approaching better the Casimir scaling.  Moreover, in \cite{ksl} the L\"{u}scher term has been calculated and found independent of the N-ality. In \cite{cglpl} the L\"{u}scher term was also found to be independent of the string tension as expected using  D4 brane embedding, appropriate for the antisymmetric representation. In the Maldacena-Nunez(MN) background \cite{mn}, which is a realization of Chamesddine-Volkov solution \cite{cvo}, by using the D3 branes an exact sine law was obtained \cite{ksk}. However, using the D5 brane embedding either in MN or Maldacena-Nastase (MNa)\cite{mna} theory, the string tension approximates a Casimir law \cite{zayasb}. In this theory it has been found a sine law, using D3 brane embedding appropriate for the symmetric representation.
Notice that several observables of the $k$-strings have been studied extensively in $\cN$=4 super-Yang-Mills plasma  \cite{Chernicoff:2006yp}. In \cite{univwl} it was shown that the energy of higher representation  Wilson loops in backgrounds with trivial dilaton is expressed in terms of the energy of the loop in the fundamental representation. Also, stable brane configurations similar to he $k$-string ones have been studied in detail \cite{Camino:2001at}.

Here we find a condition that when satisfied the observables of $k$-strings can be expressed as proportional to the observables of the fundamental strings and their string tension may be found from the proportionality constant. Our condition is related to a conserved quantity under T-duality $e^{-\phi} \sqrt{g}$, where the dilaton and the induced metric of the brane appears. This is not surprising since it naively states that in order to make an analogy of the DBI action to the Nambu-Goto action, the dilaton contribution should be cancelled out with contribution of the geometry in the internal space. This must happen in  the initial Dp-brane configuration, since the quantity is preserved under the T-dualities, acted on the brane. When the mapping is possible the $k$-string observables can be effectively reduced to single string observables and we derive the exact relation. An application of our generic analysis is presented on the non-trivial anisotropic Lifshitz-like theories, where our constraint is satisfied.

In a slightly different scheme, non-trivial mappings between the higher representation  Wilson loops \cite{Imeroni:2006rb} in the marginal  TsT $\beta$ deformed theories \cite{Lunin:2005jy} and same loops in the undeformed theories have been observed. The invariant condition used in our paper is vital for this mapping too.

Finally, motivated by the fact that the anisotropic spaces\footnote{The anisotropic gauge/gravity dualities attracted attention recently in an attempt to describe strongly coupled anisotropic systems by studying several observables eg. \cite{Giataganas:2012zy,Chernicoff:2012iq,Arefeva:2015rwa}, and due to some very interesting properties they have \cite{Rebhan:2011vd,Giataganas:2013hwa,Giataganas:2013zaa}. A review on anisotropic dualities with a collection of references can be found in \cite{Giataganas:2013lga}.} satisfy our condition we compute and additional quantity: the width of the $k$-string and we find logarithmic broadening \cite{Luscher:1980iy}. The problem is reduced to finding a 'modified' minimal surface of revolution. However, the anisotropies affect the total expression of the width, due to the different measures along the anisotropic directions.

The paper is organized as follows. In the section 2, we present the full setup for the $k$-string in a generic background and we explain when its observables can be expressed in terms of the fundamental string observables.  In section 3, we work in zero temperature and compute the width of the $k$-string in anisotropic theories to find the logarithmic broadening. We finalize the paper with conclusions and discussing our results.

\section{$k$-strings in generic gauge/gravity duality}

We consider a $d+1$ dimensional homogeneous space with the metric of the following form
\bea
ds^2 = g_{00}\prt{u} dx_0^2+g_{ii}\prt{u}dx_i^2  +g_{uu}\prt{u} du^2+ g_{\cX}\prt{u}\left(d\th^2+s_\th^2 d\cX_q^2\right)\, ,
 \label{metricgen1}
\eea
where $i=1,...,d$. The internal space consists the coordinate $\th$ and the manifold $\cX$. All the functions that appear in the metric elements depend on the holographic direction $u$. The background has also a non-trivial dilaton and a $\prt{q+1}$-form. The space we consider is  generic and includes the anisotropic backgrounds. The generalization to other space dimensions with Dp-brane probes is straightforward.

In the dual background we compute the $k$-strings in full generality and show that it is possible to express several of their observables in terms of the observables of fundamental strings in the same background.

\subsection{The probe Dp-brane Analysis}\la{sect:ft1}

The following analysis can be done with simple modifications to
an arbitrary dimensional space and for almost any arbitrary dimension of embedding brane. We will comment further on that later. To simplify the presentation we choose an internal space of 5-dimensions, and a D5-brane describing the heavy bound state. The configuration of the $k$-string is a brane which covers only two dimensions in the external space, since it consists in a sense of $k$-interacting Wilson lines. These are the time and a spatial dimension, where the static gauge is used.
It also wraps the $X^4$ of the internal space. We initially allow to the brane to have a profile along the angle $\theta$. The magnitude of $\th$ will determine the size of the $X^4$.  Since we are dealing with curved spaces we allow the profile of the brane to extend along the radial direction, so when attempting to minimize its area enters to the bulk. Therefore a suitable parametrization is
\be\la{str2a}
x_0=\t~,\quad x_1=\s~,\quad u=u(\s)~,\quad\th=\th(\s)\quad \mbox{and wrapping along the $X^4$}~,
\ee
where the $x_1$ direction is chosen to align the brane in the external space. In the case of isotropic backgrounds all directions are equivalent, but in anisotropic backgrounds the direction where the heavy probes are placed changes the energy of the configuration.

The DBI action with Wess-Zumino  term takes the form
\be\label{dbi}
S=T_{D_5}\int d\t d^5\s e^{-\phi}\sqrt{g+2  \pi \a' F_{\m\n}}-i g_{st} T_{D_5}\int 2 \pi \a'F_{\m\n}\wedge C_4~,
\ee
where
\be
\l=g_s N~,\quad T_{D_5}=\frac{N\sqrt{\l}}{8 \pi^4}~.
\ee
We write the $C_4$ form in terms of a function $D$ as
\bea
C_4:=-\frac{D\prt{\th}}{g_{st}}vol_X~,
\eea
where $vol_X$ is the volume form of the $X$ space and we rescale the electric flux as
\be
F_{\t\s}=i F\frac{1}{2\pi \a'}~.
\ee
We impose to the brane a charge quantization which becomes
\be\la{quant}
\frac{\d S}{\d F_{\t\s}}=i k\Rightarrow \frac{\d S}{\d F}=-\frac{k}{2\pi \a'}~.
\ee
The action \eq{dbi} becomes
\be\la{static1}
S=\frac{N \sqrt{\l}Vol_X}{8 \pi^4}\int d\t d\s \left(s_\th^4 h\prt{u}\sqrt{G_s-F^2}- D(\th) F\right)~,
\ee
where $G_s$ is the induced metric determinant of a fundamental string configuration, with its two-dimensional world-sheet to be parametrized by the four  functions in \eq{str2a}, and is given by
\be
G_s=g_{00}\left(g_{11}+g_{uu}u'^2+g_{\th\th}\th'^2\right)~.
\ee
The function $h\prt{u}$ is defined as
\be
h\prt{u}:=e^{-\phi}  g_{\cX}^2~.
\ee
The equations of motion for the $\th$ and the $F$ respectively give
\bea\label{eomth}
4 s_\th^3 c_\th h \sqrt{G_s-F^2}-F \partial_\th D  =\partial_1\left(\frac{s_\th^4 G_{00}G_{\th\th}\th'}{\sqrt{G_s-F^2}}\right)\\
\sqrt{G_s-F^2}=\frac{s^4_\th h F}{- D+\kt}~,\quad\mbox{where}\quad\kt:=\frac{4\pi^3 k}{N Vol_X }~.\label{eomf}
\eea
Substituting the equation \eq{eomf} to the equation \eq{eomth} we get
\be\label{eomth2}
\left(\frac{4 s_\th^7 c_\th h^2}{\kt-D}-\partial_\th D\right)F=\partial_1\left(G_{00}G_{\th\th}\frac{\kt- D}{F}\th'\right)~.
\ee
To express the energy of the Dp-brane in terms of the fundamental string we require that the equation of motion for $\th$ \eq{eomth} is satisfied for constant angle $\th=\th_0$. The equation \eq{eomth2} becomes
\be\la{eomth3}
4 h^2 c_\th s_\th^7+ D \partial_\th D- \kt \partial_\th D =0
\ee
and has non-trivial solutions for $\th=\th_0$  when the function $h\prt{u}$ satisfies
\be\la{dil1}
e^{-\phi\prt{u}} g_\cX\prt{u}^2 =c~,
\ee
which implies a cancellation in the action between the dilaton and the function $g_\cX\prt{u}^2$ of the metric.  Notice that the expression of \eq{dil1}, is the $u$ dependent part of the function $e^{-\phi} \sqrt{g}$, with $g$ being the induced metric on internal space, which has been proved to be conserved under the T-duality. If the condition \eq{dil1} is satisfied then the Dp-brane describing the high representation Wilson loop can be thought as of a string world-sheet describing the Wilson loop in the fundamental representation. We point out that this is an analogy and not a mapping in the strict sense, and can be found by acting on the brane a series of T-dualities and requiring the condition \eq{dil1} to end up with a action proportional to the Nambu-Goto. To translate the condition \eq{dil1} to a Dp-brane of different dimensions, it is more convenient to isolate the relevant part of $e^{-\phi} \sqrt{g}$ in the space under consideration. For example, this is condition satisfied for the probe D4-branes considered in the D4-brane background in \cite{Callan:1999zf}.

We set the constant $c=1$ for convenience, and from now on we have the angle $\th$ to be a constant depending on $\kt$, specified by the solution of equations \eq{eomth3} with the constrain \eq{dil1}. This relation is not as constraining as it seems and it is satisfied in several backgrounds.

The equation \eq{eomf} can be solved for $F$ and the solution may be written as
\be\la{dez}
F=\pm\frac{\sqrt{G_s}}{\sqrt{1-\frac{s_{\th} \partial_{\th} D}{4  c_{\th}}}}:=\pm Z\prt{\th}\sqrt{G_s}~,
\ee
where we choose the positive sign. The above equality plays a key role in expressing the whole DBI action in terms of the Nambu-Goto. The action of the D5-brane becomes
\be
S_{D5}=\frac{N\sqrt{\l} Vol_\cX}{8 \pi^4} s_\th^4 \int_{u_0}^{u_b}d\t d\s\prt{\sqrt{1-Z\prt{\th}^2}- Z\prt{\th} D(\th)}  \sqrt{G_s}~,
\ee
where the integration is done from the boundary to the turning point for the surface and the infinities have not yet subtracted with the counterterms.

At this stage one may also like to impose the N-ality, which would require that the redefinition $k\rightarrow N-k$
implies a symmetric transformation in the function $G\prt{\th}$ defined as
\be\la{eomth3a}
G\prt{\th}:= \frac{\prt{4 h^2 c_\th s_\th^7+D \partial_\th D} Vol_\cX}{4 \pi^3 \partial_\th D}=\frac{k}{N}
\ee
where we have used \eq{eomth3}. Such a possible transformation is that for $k\rightarrow N-k$, which could result for the function $G\prt{\pi-\th}=G\prt{\th}-1$.

Due the way that the Dirichlet and Neumann boundary conditions are imposed to the brane at the boundary, the on-shell action will have divergences which need to be canceled with a Legendre transform. To apply this method for the successful cancelation of the divergences, one need to make sure that the background satisfies asymptotically the conditions obtained in \cite{giataganasUV}. In the case of the $AdS\times S$ background the method has been applied for strings and D-branes in \cite{DGO,wldd}.

In our case we assume that the conditions of \cite{giataganasUV} are satisfied and the boundary terms that need to be added in action \eq{dbi} are
\be\la{bound1}
S_{D_5~b}=-\int_\partial d\s x^\m\frac{\d S_{D_5}}{\d\partial_\s x^\m}-\int_\partial d\s \frac{\d S_{D_5}}{\d\partial_\s \th}+\frac{\sqrt{\l}}{2 \pi}\int d\t d\s k F~.
\ee
The first term, where $x^\m$ should be taken as the radial direction, regularizes the action of the fundamental string configuration and can be thought as coming from the Nambu-Goto action. The second term of the boundary action is zero for constant $\th$.
Therefore to compute the total action we need to take into account the on-shell action \eq{dbi} and the remaining boundary term from \eq{bound1} which gives the compact result
\be\la{final1}
S_{D5}
= \frac{N Vol_\cX}{4 \pi^3} \frac{s_\th^4}{ \sqrt{1-Z\prt{\th}^2}} S_{NG,normalized}~.
\ee
To obtain \eq{final1} we have solved the conditions \eq{quant} for $k$ and substituted to the total action. We remind that the constraint
\eq{dil1} requiring cancelation of the dilaton dependence in the action with an overall term of the induced geometry is crucial in order to
express the k-string in terms of the fundamental string.
The minimization of the Nambu-Goto action for a orthogonal Wilson loop in a generic background can be found for example in \cite{stringtension,Giataganas:2012zy}. Therefore the expression  \eq{final1} is directly applicable to any background satisfying the condition \eq{dil1}.

The fact that the condition is satisfied for many backgrounds can be understood by several qualitative arguments.
We have mentioned already one with the T-dualities and the preserved quantity. Moreover, the k-string as a bound state from a distance would look as a compact string, if we ignore its structure, and it would be natural to expect that its properties can be expressed in terms of the Nambu-Goto action.

\subsection{Drag Force on k-Quarks and Gluons}

In this section we briefly comment on other observables of $k$-strings and gluons. These can be expressed in terms of corresponding observables of the fundamental strings, as a straightforward generalization of the static computation.  The situation is analogous to that of the k-strings in the $AdS_5\times S^5$ space \cite{Chernicoff:2006yp}.

To compute the dragging of the $k$-quarks, let us parametrize the $k$-string that move with a constant speed $v$ along the $x_1$ direction with the following ansatz
\be
x_0=\t~,\quad u=\s~,\quad \th=\th\prt{\s}~,\quad x_1\prt{t,\s}=v t+\xi\prt{\s}~.
\ee
The resulting action is similar to the stating configuration \eq{static1} but the induced metric  $G_s$ is given by
\be
G_s=\prt{g_{00}+v^2 g_{11}}\prt{g_{uu}+\th'^2 g_{\th\th}}+\xi'^2 g_{00} g_{11}~.
\ee
We require that in the dual background the contribution of the dilaton in the action cancels with the induced geometry, ie.
the condition \eq{dil1} is satisfied. The drag force of the k-string is calculated by the momentum flowing from the boundary to the horizon of the space and turns out to be
\be
F_k=\frac{N\sqrt{\l} Vol_\cX}{8 \pi^4} \frac{s_\th^4}{ \sqrt{1-Z\prt{\th}^2}}
 F_{string}~,\quad\mbox{where}\quad F_{string}=\frac{1}{2 \pi\a'}\sqrt{g_{tt} g_{11}}|_{u=u_0}
\ee
where the angle $\th$ is chosen to satisfy \eq{eomth3} and $Z$ is given by \eq{dez}. The radial distance $u_0$ is the horizon of the induced world-sheet metric given by solving the equation\footnote{A generic treatment of observables in the gravity duality may be found in \cite{Giataganas:2012zy}.}
\be
g_{uu}\prt{g_{00}+g_{11} v^2}=0~.
\ee
The drag force of k-quarks is proportional
to the single quark dragging with proportionality factor
\be
a(\th):=\frac{s_\th^4}{ \sqrt{1-Z\prt{\th}^2}}~.
\ee
Some interesting remarks are in order that reveal the properties of $a\prt{\th}$. We expect that the number of quarks in the bound state that maximize the drag force is $k=N/2$ since for this value the color charge of the string configuration is maximum. For $k=0$ or $k=N$ the drag force becomes zero and that could correspond to the conical part of the D5-brane becomes minimum and zero. It is  a $k$-string that is color neutral,  with no quarks, or N number of quarks making a baryon. This could happen at $\th=0$ or $\th=\pi$ unless the overall sine dependence of the numerator of $a(\th)$ is cancelled.
Notice that another way to determine the expected maximum of dragging is to think that the two minima are for $k=0, N$ and the maximum is expected in the middle for symmetry reasons of the N-ality. Moreover, we expect that the k-string, say for $k=N/2$, experience less dragging than a number of $N/2$ single quarks which experience $k F_1$ drag force. This also constrains the properties of the function $a(\th)$.

Let us also comment on the gluons dragging. We may think the gluon as two parallel D5-branes carrying $k=1$ and $k=N-1$ units of fundamental string charge \cite{Gomis:2006sb,Chernicoff:2006yp}. A proposal for the action is given by \cite{Myers:1999ps}, where for generic diagonal backgrounds, with the additional cancelation of the dilaton contributions \eq{dil1}, the dynamics of the two D5-branes decouple each other at leading orders. By repeating the analysis of the $k$-strings we get that the force on the gluon is twice the force on the single quark. This is because the energy of the $k=N-1$ k-string is equal to the force on the $k=1$ due to the N-ality.
\be
F_{drag,gluon}=2 F_{drag,quark}~.
\ee
\newline
\textbf{Application to a particular background:}\newline
As a simple non-trivial example we look at the anisotropic supergravity background which is a deformed version of the $\cN=4$ finite temperature sYM \cite{mateosaniso}, presented briefly here. In the dual field theory a $\th$-parameter term is present, which depends on the anisotropic direction and is related to the axion of the type IIB supergravity through the complexified coupling constant of the $\cN=4$ sYM. In the gravity dual background the anisotropy is generated due to existence of axion term depending on the anisotropic direction. In the string frame the background is given by
\bea
&&\hskip -.20cm
ds^2 =
 \frac{1}{u^2}\left( -\cF \cB\, dx_0^2+dx_1^2+dx_2^2+\cH dx_3^2 +\frac{ du^2}{\cF}\right)+ {\cal H}^{-\frac{1}{2} \, }\left(d\th^2+s_\th^2 d\Omega_4^2\right)\,,
 \label{metric111} \\
&& \hskip -.20cm \chi = a x_3, \qquad \phi=\phi(u) \, ,
\label{chi1111}
\eea
where $a$ is the anisotropic parameter with units of inverse length,  $\phi$ is the dilaton and $\chi$ is the axion.  For our purposes is also important a RR five form which reads
\be
F_5=  \left(\Omega_{S^5}+\star\Omega_{S^5}\right)~.
\ee
In the small anisotropy over temperature limit $T\gg a$, the metric can be found analytically
\bea
\cF(u) &=& 1 - \frac{u^4}{u_h^4} + a^2 \cF_2 (u)  +\mathcal{O}(a^4)~,\quad
\cB(u) =1 + a^2 \cB_2 (u) +\mathcal{O}(a^4)\,, \\
\cH(u)&=&e^{-\phi(u)},\quad\mbox{where}\quad \phi(u) =  a^2 \phi_2 (u)  +\mathcal{O}(a^4)\, ,
\label{smallae}
\eea
where the form of functions is given in \cite{mateosaniso}. Notice that the $u$ dependent part of the $e^{-\phi} \sqrt{g}=1$, so the condition \eq{dil1} is satisfied.  Therefore, the $k$-string energy and observables on this background follow exactly the results of the single string observables of \cite{Giataganas:2012zy}.

\section{Width of the k-strings in general theories}

In this section we study the width of the k-string at the zero temperature using the generic metric \eq{metricgen1} and then we apply our formulas to the interesting for this case anisotropic Lifshitz-like backgrounds.

Due to the rotational symmetry of our configuration we make a coordinate transformation in the initial metric \eq{metricgen1} to write it in the form
\be
ds^2=g_{rr}\prt{dr^2+r^2 d\phi^2}+ g_{zz} dz^2+g_{jj} dx_j^2+ g_{uu}du^2 + g_{\cX}\prt{u}\left(d\th^2+s_\th^2 d\cX_4^2\right)\, ,
\ee
where all the metric elements are functions of the holographic coordinate $u$.
To compute the width of the $k$-string we need we consider the catenoid brane configuration ending on two different radii $R_1>R_2$ on the boundary given by the following parametrization
\be
r=r\prt{\s}~,\quad \phi\prt{\t}=\t~,\quad z\prt{\s}=\s~,\quad \th\prt{\s}=\s ~,
\ee
with the boundary conditions
\be
r\prt{0}=R_1~,\quad r\prt{L}=R_2~.
\ee
The theories under investigation are in principle not confining, therefore in order to compute the width of the $k$-string, we introduce a hard-wall version of the model and compute the observables at $u=u_\L$.

Using the DBI action \eq{dbi} we get
\be\la{dbi1}
S=\frac{N\sqrt{\l} Vol_\cX}{8 \pi^4} \int d\t d\s  s_\th^4 g_\cX^2 e^{-\phi} r \sqrt{g_{rr}\prt{r'^2g_{rr}+ g_{zz}+g_{\th\th}\th'^2}- F^2} + D\prt{\theta} F~.
\ee
We require the condition for the dilaton \eq{dil1} to be satisfied and the angle $\th$ is a constant given by the equation \eq{eomth3a} with
\be
G_s=r \sqrt{g_{rr}\prt{r'^2g_{rr}+ g_{zz}}}~.
\ee
The total action is expressed again in terms of the fundamental string
\be
S_{D5}=\frac{N\sqrt{\l} Vol_\cX}{8 \pi^4} \frac{s_\th^4}{ \sqrt{1-Z\prt{\th}^2}} \int_{0}^{L}d\t d\s \sqrt{G_s}~.
\ee
The problem effectively is now reduced on finding the width of the world-sheet in an anisotropic space.  The equations of motion of \eq{dbi1} are
\bea\la{emswidth}
g_{rr} \sqrt{r'^2  +\frac{g_{zz}}{g_{rr} }}-\partial_1\prt{\frac{r' r g_{rr}}{\sqrt{r'^2  +\frac{g_{zz}}{g_{rr} }}}}=0&&~,\\
\frac{r g_{zz} }{\sqrt{r'^2+\frac{g_{zz}}{g_{rr}}}}=c&&~.
\eea
Combining them we end up with the simplified equation where its solution specifies the profile of the tube-like brane
\be
r'' r-r'^2+\frac{g_{zz}}{g_{rr}}=0~.
\ee
It is a simple differential equation even when anisotropies are present. By constraining the surface at the radial distance $u_\L$ to introduce the additional scale in the theory, we have reduced the problem to be effectively flat. In the isotropic case, the last term is equal to the unit and the solutions of the equation correspond to a deformed minimal surface of revolution.

An analytical solution of a compact form may be found
\be
r= c\sqrt{\frac{g_{zz}}{g_{rr}}} \cosh\prt{\frac{\s-\s_0}{c}}~,
\ee
with $\s_0$ the point where the radius of the brane becomes minimum, satisfying $r'\prt{\s_0}=0$. Applying the boundary conditions we get
\be\la{bbb1}
R_1= c\sqrt{\frac{g_{zz}}{g_{rr}}} \cosh\frac{\s_0}{c}~,\quad R_2= c\sqrt{\frac{g_{zz}}{g_{rr}}} \cosh\frac{L-\s_0}{c}~,
\ee
where the constant $c$ cannot be expressed analytically in terms of the bound state sizes of the  loops. We have taken $R_1>R_2$ and this is in agreement with \eq{bbb1} since $\s_0>L-\s_0$ and  the minimum of the radius is closer to the small radius circle.
The on-shell action becomes
\be
S=\frac{N\sqrt{\l} Vol_\cX}{8 \pi^4} \frac{s_\th^4}{ \sqrt{1-Z\prt{\th}^2}} g_{zz} c \int d\s d\t \cosh^2\frac{\s-\s_0}{c}~,
\ee
which can be integrated to give
\be
S=\frac{N\sqrt{\l} Vol_\cX}{8\pi^3} \frac{s_\th^4}{ \sqrt{1-Z\prt{\th}^2}}c  g_{zz} \prt{L+\frac{c}{2}\prt{\sinh\frac{2 \prt{L-\s_0}}{B}}+\sinh\frac{2 \s_0}{B}}~.
\ee
In the limit of large radius $R_1$ specified as $R_1\gg 1$ and $R_1\gg R_2$ the constant $c$ in leading order is solved analytically giving
\be
c\simeq\frac{L}{\log\frac{R_1}{R_2}}~,
\ee
where the  subleading terms have been ignored. Notice that the constant does not depend on the metric elements.
The action takes the final form
\be
S=\frac{N\sqrt{\l} Vol_\cX}{8\pi^3} \frac{s_\th^4}{ \sqrt{1-Z\prt{\th}^2}}\prt{g_{zz}
\frac{L^2}{\log\frac{R_1}{R_2}} +g_{rr}  \prt{R_2^2-R_1^2}}~,\la{enfin1}
\ee
and up to different constants multiplying the 'disk area' term depending on the radii and the $L^2$ term, is similar to the one of
the isotropic flat space, while for $g_{zz}=g_{rr}$ reduces to it.
The width of the bound state measures how thick is the tube, by measuring the chromoelectric flux through a probe Wilson loop. It is defined as \cite{Luscher:1980iy}
\be
w^2:=\frac{\int e^{-S}L^2dL}{\int e^{-S}dL}~.
\ee
The $L$-independent terms of \eq{enfin1} cancel and to the leading order we get the width
\be\la{width11}
w^2=\frac{4\pi^3\a'}{N Vol_\cX} \frac{\sqrt{1-Z\prt{\th}^2}}{ s^4_\th g_{zz}}\log{\frac{R_1}{R_2}}~.
\ee
The logarithmic broadening of the k-string is present in the case of an anisotropic space. However an essential difference with the isotropic result is the form of the proportionality factor and the string tension. The width of a fundamental string is known to be
\be
w^2=\frac{1}{2 \pi\s}\log{\frac{R_1}{R_2}}~.
\ee
The k-string tension can be identified as the proportionality factor from the equation \eq{width11}.
In the isotropic space this is straightforward, since $\s=g_{rr}(u_\L)=g_{zz}(u_\L)$ and the string tension of a fundamental
string is the same in along all the directions. In an anisotropic space, the string tension is different along the different directions.

To understand better our result \eq{width11}, let us point out that in an accurate treatment the original configuration of the $k$-string considered here, should ideally consist of a bound state
of multiple quarks and antiquarks lying along the spatial $x_1=r \cos \phi$ direction with the size of the state being $R_0$.
This would form an orthogonal configuration at the boundary where the minimal surface is attached. However, for the
computation of the width of the bound state, in the limits we are working, we have assumed a simplified version of the orthogonal configuration, to be cyclic of radius $R_1\simeq R_0$ in order to
simplify the computation. This approximation does not affect the logarithmic broadening \cite{Luscher:1980iy}. Therefore,
in the anisotropic space the width of a single heavy meson bound state along the $x_1$ direction with metric element $g_{rr}$, could be written as
\be
w^2=\frac{g_{rr}}{2 \pi \s g_{zz}}\log{\frac{R_1}{R_2}}\quad \mbox{with}\quad \s=g_{rr}(u_\L)~.
\ee
Then the width of a $k$-string along the $x_1$ direction
\be\la{widthf1}
w^2=\frac{g_{rr}}{2 \pi \s_k g_{zz}}\log{\frac{R_1}{R_2}}\quad \mbox{with}\quad \s_k=\frac{N Vol_\cX}{2\pi^2\a'} \frac{ s^4_\th g_{rr}(u_\L)}{\sqrt{1-Z\prt{\th}^2}}~.
\ee
Our formulas are valid for the isotropic case, and reproduce the well known results. In the anisotropic space the result is sensitive to the way we measure the width, due to the fact that the
distance that the probe Wilson loop of radius $R_2$ is separated from the fundamental state has different measure than the direction
along where the bound state is aligned. Therefore, we propose that an appropriate definition of the width in anisotropic spaces
is given by the following expression
\be\la{widthf12}
w_{aniso}^2:=\frac{g_{zz}}{g_{rr}}w^2 ~,
\ee
which eliminates the dependence of the width of the bound state from the test 'measuring' Wilson loop.

Let us apply the analysis to an anisotropic zero temperature Lifshitz-like solution found in \cite{Azeyanagi:2009pr} with metric
\be
ds^2=\tilde{R}_s^2 \prt{u^{7/3}\prt{-dt^2+dx_1^2+dx_2^2}+u^{5/3}dz^2+ \frac{1}{u^{5/3}}du^2}+R_s^2 u^{1/3} d^2s_{S_5}~
\ee
 dilaton
\be
e^\phi=u^{2/3} e^{\phi_0}~,
\ee
and a 5-form equal to that of $AdS_5\times S^5$.
Notice that $e^{-\phi} g_{S^5}^2=1$ and the dilaton contribution to the action is canceled by the relevant part of the geometry.
By applying the formula \eq{widthf1} the width of the anisotropic $k$-string is given by
\be
w^2_{aniso}= u_\L^{-2/3} w^2=\frac{1}{2 \pi\s_k }\log{\frac{R_1}{R_2}}=\frac{3 \pi \a'}{2  N s_\th^3 u_\L^{7/3} }\log{\frac{R_1}{R_2}}~.
\ee

\section{Conclusions}

We have studied the properties of $k$-strings in a generic gauge/gravity duality. We have found that when the dilaton contribution cancel with the geometry of the induced metric in the Dp-brane action, then the observables of the $k$-strings are proportional of the observables of fundamental strings in the same space. The observables include the energy of the $k$-string bound state and the dragging of the moving $k$-string. The extension of our work to other space dimensions and probe brane dimensions, as well observables of similar type should be straightforward.

The condition \eq{dil1} is part of an invariant quantity under a T-duality. This is not surprising since under the T-dualities acting on the Dp-brane, the resulting two-dimensional action, can be brought to a form proportional to the Nambu-Goto action only if the equation \eq{dil1} is satisfied. We point out that strictly speaking this is an analogy between the Dp-brane and the string actions, at the level of equations of motion and not a strict mapping. However, when the conditions derived are satisfied it can be naively thought as a mapping.  Moreover, one may understand our result as examining the $k$-string from a distance, where its microscopic structure is not visible. It is natural for the multi-quark bound state to look as a single string with a tension $\s_k$. It is easy to identify the $k$-string tension, of Casimir, sine or other type, from the way that it is related to the fundamental string tension.

Notice that even in finite temperature when the condition \eq{dil1} is satisfied the energy of the $k$-string is written in the form \eq{final1}, where the induced metric of the string is now on the black hole background. It would be interesting however to examine how the string tension of the $k$-string is affected in the finite low temperature confining phase. The string models predict a decrease of the potential of the heavy meson state where the string tension decreases as temperature increases  \cite{Pisarski:1982cn, deForcrand:1984cz, Gao:1989kg}. This has been noticed recently using also the gauge/gravity duality \cite{Giataganas:2014mla}. In finite temperature but still confining phase we expect that it would not be possible to express the $k$-string energy in terms of the single string.

As a side remark and an interesting further direction we point out that the invariance of the condition \eq{dil1} has been also observed in the non-Abelian T-duality as described in \cite{Itsios:2013wd,Macpherson:2014eza}. In these backgrounds one may think in a bottom-up way to propose the description of the high representation Wilson loops with probe Dp-branes that satisfy our condition.

We notice that the condition \eq{dil1} is satisfied for the anisotropic top-down supergravity solution. Motivated by the fact that the $k$-strings in these spaces can be written as fundamental strings we have computed the width of this state. The width of gluonic string of the heavy meson as well as of a $k$-string state has been found to have a logarithmic broadening in the certain confining theories \cite{Luscher:1980iy,Giudice:2006hw,Greensite:2000cs,Armoni:2008sy}, which is a property of the heavy meson bound states in all confining theories \cite{Giataganas:2015yaa}. The $k$ string bound state is placed on a plane with rotational symmetry, while the probe Wilson loop is placed at a distance from the $k$-string loop in the anisotropic direction. The width has a logarithmic broadening as in the isotropic space, however the proportionality factors are affected by the anisotropy of the space. We argue that this is due to different measures along the different directions, and we propose a slightly modified definition of the width in the anisotropic space.
As a further step to this computation it would be interesting to place the $k$-string brane on a plane that is anisotropic.

\startappendix

\textbf{Acknowledgements:} We are thankful to A. Armoni, A. Guijosa, C. Nunez, A. Ramallo, K. Sfetsos and M. Teper, for useful conversations and comments. The research of D.G. is partly supported by a Fellowship of State Scholarships Foundation, through the funds of the “operational programme education and lifelong learning” by the European Social Fund (ESF) of National Strategic Reference Framework (NSRF) 2007-2013.


\end{document}